\documentclass[]{jpsj3}
\usepackage{graphicx}
\usepackage{amsmath}
\usepackage[dvips]{color}

\title{Roles of Potential Gradient and Electrode Bandwidth on Negative Differential Resistance 
in One-Dimensional Band Insulator
}

\author{
Yasuhiro Tanaka$^{1,2\ast}$
and Kenji Yonemitsu$^{1,2}$}

\inst{
$^{1}$Department of Physics, Chuo University, Bunkyo, Tokyo 112-8551, Japan\\
$^{2}$JST, CREST, Chiyoda, Tokyo 102-0076, Japan
}

\date{\today}

\abst{
A negative differential resistance (NDR) in a one-dimensional band insulator attached 
to electrodes is investigated. We systematically examine the effects of an electrode bandwidth and 
a potential distribution inside the insulator on current-voltage characteristics. We show 
that, in uncorrelated systems, the NDR is generally caused by a linear potential gradient as well as 
by a finite electrode bandwidth. In particular, the former reduces the effective bandwidth of the 
insulator for elastic tunneling by tilting its energy band, so that it brings about 
the NDR even in the limit of large electrode bandwidth.
}

\begin{document}
\sloppy
\maketitle

\section{Introduction}
Recently, nonlinear conduction phenomena in low-dimensional electron systems 
such as Mott insulators\cite{Taguchi_PRB00,Yamanouchi_PRL99} and charge-ordered 
materials\cite{Mori_JMC07} have been intensively studied. In theoretical 
investigations, several authors consider a model structure 
where an insulator with a length 
$L_C$ in the central part is attached to the left and right ($\alpha =L, R$) 
electrodes\cite{Yonemitsu_JPSJ05,Oka_PRL05b,Yonemitsu_PRB07,Yonemitsu_JPSJ09,Okamoto_PRB07,
Okamoto_PRL08,Ajisaka_PTP09,Heidrich_PRB10,Tanaka_PRB11,Tanaka_JPSJ11,Tanaka_JPC13}. 
A schematic picture of the model for a one-dimensional case is shown in Fig. \ref{fig:fig1}. 
When a bias voltage is applied, the electrode bandwidth and the electrode density of states 
are important factors for determining current-voltage ($J$-$V$) characteristics. 
In fact, the finite electrode bandwidth results in a negative differential 
resistance (NDR) if we consider only elastic electron tunneling for transport. 
This NDR, which is not related to the electron correlation, has been shown using a 
noninteracting resonant level model\cite{Baldea_PRB10} in which the central part 
consists of a single site ($L_C=1$). 
When $L_C$ becomes large, the potential distribution inside the 
central part is also important. It gives a spatial 
dependence of the electric field and affects the breakdown mechanism of 
insulators\cite{Tanaka_PRB11}. 
Although it is expected that the NDR depends on both 
the potential distribution and the electrode degrees of freedom, 
their effects have not been examined so far for large-$L_C$ systems. 

\section{Model and Method}
\begin{figure}
\begin{center}
\includegraphics[height=2.0cm]{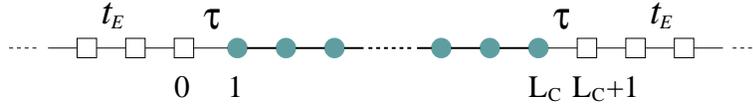}
\end{center}
\caption{(Color online) Schematic structure of the model. Left and right semi-infinite electrodes 
are connected by a one-dimensional band insulator. Solid (open) symbols 
represent the sites in the central part (electrodes).}
\label{fig:fig1}
\end{figure}

In this paper, we study the $J$-$V$ characteristics of a one-dimensional band 
insulator at half-filling, to which semi-infinite electrodes are attached (Fig. \ref{fig:fig1}). 
We assume that electrons are noninteracting in both the central part and the electrodes. 
Although we do not consider insulators that are caused by electron-electron interactions here, 
some of our results will be qualitatively applied to such insulators at least on the mean-field 
level\cite{Tanaka_PRB11}. 
The total Hamiltonian is written as $H=H_L+H_R+H_C+H_{con}$ with
\begin{eqnarray}
H_L&=&-t_E\sum_{i\leq 0, \sigma}(c^{\dagger}_{i\sigma}c_{i-1\sigma}+h.c.)+\mu_L\sum_{i\leq 0, \sigma}
c^{\dagger}_{i\sigma}c_{i\sigma},\\
H_R&=&-t_E\sum_{i\geq L_C+1, \sigma}(c^{\dagger}_{i+1\sigma}c_{i\sigma}+h.c.)+\mu_R\sum_{i\geq L_C+1, \sigma}
c^{\dagger}_{i\sigma}c_{i\sigma},\\
H_C&=&-\sum_{i=1, \sigma}^{L_C-1}[t+(-1)^{i+1}\delta t](c^{\dagger}_{i+1\sigma}c_{i\sigma}+h.c.)+
\sum_{i=1, \sigma}^{L_C}V_C(i)c^{\dagger}_{i\sigma}c_{i\sigma}\label{eqn:Hc},\\
H_{con}&=&-\tau \sum_{\sigma}(c^{\dagger}_{0\sigma}c_{1\sigma}+c^{\dagger}_{L_C\sigma}c_{L_C+1\sigma}+h.c.), 
\end{eqnarray}
where $H_{\alpha}$ and $H_C$ are the Hamiltonians for the electrode $\alpha$  
and the central part, 
respectively, and $H_{con}$ describes the coupling between them. 
$c^{\dagger}_{i\sigma}$ ($c_{i\sigma}$) denotes the creation (annihilation) operator 
for an electron with spin $\sigma$ at the $i$th site, and ${\it h.c.}$ stands for 
hermitian conjugate. 
For the electrodes, the transfer integral is denoted by $t_E$ and their coupling to the 
central part is denoted by $\tau$. 
In Eq. (\ref{eqn:Hc}), the transfer integral $t$ has a modulation $\delta t$ that gives a 
charge gap $\Delta =4\delta t$. We use $t$ as a unit of energy. 
In the following, the bandwidth of the central part (the electrodes) is written as 
$W$ ($W_E$) where $W\simeq 4t$ if $t\gg \delta t$ holds, and $W_E=4t_E$. 
All these quantities are summarized in Table I. 
Since we do not consider the work-function difference at the interfaces, we set $\mu_L=V/2$ ($\mu_R=-V/2$)
 for the chemical potential of the electrode $L$ ($R$) when the bias $V$ is applied. 
$V_C(i)$ is the site potential for which we consider two cases. They are written as
\begin{eqnarray}
V_C(i)=V[1/2-i/(L_C+1)],\label{eq:uni_E}
\end{eqnarray}
and
\begin{eqnarray}
V_C(i)=0\label{eq:no_E}. 
\end{eqnarray}
In Eq. (\ref{eq:uni_E}), the central part has a linear potential gradient that corresponds to a 
uniform electric field; however, in the case of Eq. (\ref{eq:no_E}), there is no electric field: 
the voltage drop occurs only at the interfaces. 

\begin{table}
\caption{Quantities that characterize our model. The independent parameters are $\delta t$, 
$t_E$, and $\tau$.}
\begin{tabular}{c|cc}\hline
Location & Quantity & Meaning\\ \hline
& $t$ ($=1$) & Transfer integral\\ \cline{2-3}
Central part & $\Delta$ ($=4\delta t$) & Charge gap\\ \cline{2-3}
& $W$ & Bandwidth\\ \hline 
Electrodes & $t_E$ & Transfer integral\\ \cline{2-3}
& $W_E$ & Bandwidth\\ \hline
Interface & $\tau$ & Transfer integral between the central part and the electrodes\\ \hline
\end{tabular}
\end{table}

\begin{table}
\caption{Conditions for calculations (i)-(iv) and calculation results for NDR.}
\begin{tabular}{cccc}\hline
Condition & Potential gradient & Electrode bandwidth & NDR\\ \hline
(i) & Present & Finite & Present\\ \hline
(ii) & Present & Infinite & Present\\ \hline
(iii) & Absent & Finite & Present\\ \hline
(iv) & Absent & Infinite & Absent\\ \hline
\end{tabular}
\end{table}

\begin{figure}
\begin{center}
\includegraphics[height=4.0cm]{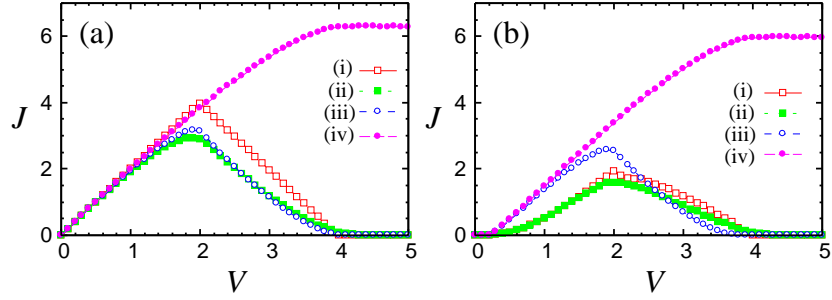}
\end{center}
\caption{(Color online) $J$-$V$ characteristics with (a) $\delta t=0$ and (b) $\delta t=0.05$ 
obtained under the conditions (i)-(iv) that are described in the text and summarized in 
Table II. We use $t_E=\tau=1$ and $L_C=100$.}
\label{fig:fig2}
\end{figure}

We use a nonequilibrium Green's function method\cite{Haug_BOOK} in order to calculate the 
$J$-$V$ curve, which can be carried out exactly since there are no electron-electron 
interactions in $H$. The retarded Green's function for the central part is written as
\begin{eqnarray}
  G^{r}(\epsilon)=[\epsilon {\bf 1}-H_C-\Sigma^r_L(\epsilon)-\Sigma^r_R(\epsilon)]^{-1}\label{eq:G},
\end{eqnarray}
where $\Sigma^r_{\alpha}(\epsilon)$ is the self-energy due to the electrode $\alpha$ and ${\bf 1}$ is 
the unit matrix. 
The spin index is abbreviated. For the self-energies, the only nonzero matrix element is given by 
\begin{eqnarray}
[\Sigma^r_{\alpha}(\epsilon)]_{i_{\alpha},i_{\alpha}}=\tau^2 [g^r_{\alpha}(\epsilon)]_{p_{\alpha},p_{\alpha}}\label{eq:Se},
\end{eqnarray}
where $g^r_{\alpha}(\epsilon)$ is the Green's function of the isolated electrode $\alpha$, 
$i_L$ ($i_R$) is the leftmost (rightmost) site in the central part, and $p_L$ ($p_R$) is the 
site adjacent to $i_L$ ($i_R$) in the electrode $L$ ($R$). 
We can obtain $[g^r_{\alpha}(\epsilon)]_{p_{\alpha},p_{\alpha}}$ as
\begin{eqnarray}
[g^r_{\alpha}(\epsilon)]_{p_{\alpha},p_{\alpha}}=
\begin{cases}
\frac{\epsilon -\mu_{\alpha}}{2t_E^2}-\frac{i}{2t_E^2}\sqrt{4t_E^2-(\epsilon -\mu_{\alpha})^2} 
& (|\epsilon -\mu_{\alpha}|<2t_E)\\
\frac{\epsilon -\mu_{\alpha}}{2t_E^2}-{\rm sgn}(\epsilon -\mu_{\alpha})\frac{1}{2t_E^2}
\sqrt{4t_E^2-(\epsilon -\mu_{\alpha})^2} & (|\epsilon -\mu_{\alpha}|>2t_E), 
\end{cases}
\end{eqnarray}
where ${\rm sgn}(x)=1$ [${\rm sgn}(x)=-1$] for $x>0$ ($x<0$) and we assume $t_E>0$. 
If we introduce $\Gamma_{\alpha}(\epsilon)=-2{\rm Im}[\Sigma^r_{\alpha}(\epsilon)]_{i_{\alpha},i_{\alpha}}$ with 
$\Gamma_{\alpha}(\epsilon)=\frac{\tau^2}{t_E^2}\sqrt{4t_E^2-(\epsilon -\mu_{\alpha})^2}
\theta (2t_E-|\epsilon-\mu_{\alpha}|)$, the current $J$ is given by\cite{Haug_BOOK}
\begin{eqnarray}
J=2\int^{\mu_L}_{\mu_R} d\epsilon \Gamma_L(\epsilon) \Gamma_R(\epsilon)|[G^r(\epsilon)]_{1,L_C}|^2\label{eq:J},
\end{eqnarray}
where we set $e=h=1$. 
When we compute $J$, an assumption that the electrode bandwidth $W_E$ is much larger 
than the other energy scales is sometimes used. This is called the wide-band limit (WBL). 
In such a case, the self-energy is independent of energy and is reduced to 
$[\Sigma^r_{\alpha}(\epsilon)]_{i_{\alpha},i_{\alpha}}=-i\Gamma_0/2$ with 
$\Gamma_0=2\tau^2/t_E$\cite{Jauho_PRB94}. 

\section{Results}
In order to elucidate the roles of the potential gradient and the electrode bandwidth, 
in the following, we use four conditions (i)-(iv) in the calculations, 
as summarized in Table II, depending on $V_C(i)$ and on whether the WBL is applied: 
$V_C(i)$ is given by 
Eq. (\ref{eq:uni_E}) [Eq. (\ref{eq:no_E})] in (i) and (ii) [(iii) and (iv)], 
and the WBL is used in (ii) and (iv). We show the $J$-$V$ curves for 
$\delta t=0$ and $\delta t=0.05$ in Figs. \ref{fig:fig2}(a) and \ref{fig:fig2}(b), 
respectively, where we use $t_E=\tau=1$ and $L_C=100$. Although the cases (i) and (ii) 
for $\delta t=0$ and the cases (iii) and (iv) for $\delta t=0.05$ involve an artificial 
voltage drop in the central part, we show their results for comparison. 
When $\delta t=0$, the central part is a noninteracting metal, so that $J$ shows a linear 
increase for small values of $V$. This feature appears regardless of the electrode bandwidth and 
$V_C(i)$. 
The slope in the linear regime is 2, which corresponds to the Landauer formula $G=\frac{2e^2}{h}T$ 
with the transmission probability $T=1$. When the central part is a band insulator 
($\delta t =0.05$), the behavior of the $J$-$V$ curves in the small-$V$ region is different from 
that for $\delta t=0$ because of the charge gap $\Delta$. 
Moreover, it depends on $V_C(i)$. In (i) and (ii), the current is described by 
$J\sim Ve^{-V_{\rm th}/V}$ indicating the Landau-Zener (LZ) 
breakdown\cite{Landau_SOW32,Zener_PRSL32} of the insulator; however, $J=0$ for $V<\Delta=0.2$ 
and it begins to increase at $V=\Delta$ in (iii) and (iv). These different behaviors depend on  
whether the electric field exists in the central part, as discussed in Ref. 12. 

\begin{figure}
\begin{center}
\includegraphics[height=6.5cm]{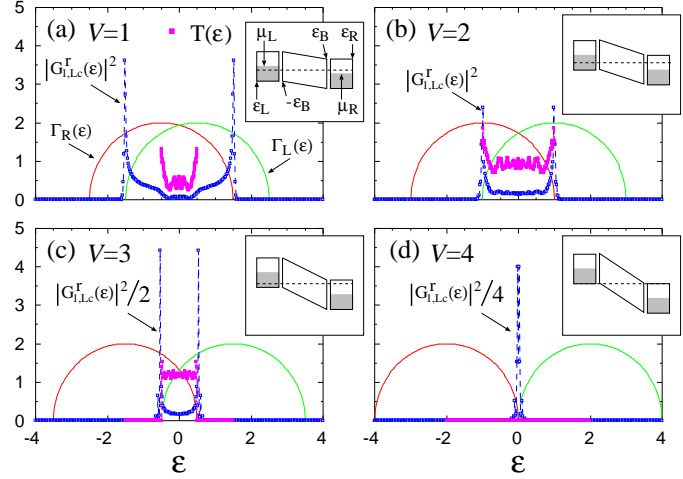}
\end{center}
\caption{(Color online) $\Gamma_L(\epsilon)$, $\Gamma_R(\epsilon)$, and $|G_{1,L_C}(\epsilon)|^2$ as a 
function of $\epsilon$ for several values of $V$ in the case of (i). 
$T(\epsilon)=\Gamma_L(\epsilon)\Gamma_R(\epsilon)|G_{1,L_C}(\epsilon)|^2$ is also shown for 
$\mu_R<\epsilon <\mu_L$. The parameters are the same as those in Fig. \ref{fig:fig2}(b). 
The inset shows schematic representations of the energy bands of the 
electrodes and the central part, where the dashed line indicates $\epsilon=0$. 
}
\label{fig:fig3}
\end{figure}

When $V$ is large, the $J$-$V$ curves for the metal and the band insulator are qualitatively 
similar. For (i)-(iii), the NDR occurs for $V\gtrsim 2$, whereas $J$ monotonically increases and 
saturates at $V\simeq 4$ in (iv). There are two different sources of these NDRs. 
One is a finite electrode bandwidth, which has been discussed in the resonant level 
model\cite{Baldea_PRB10}. When $V$ exceeds $W_E/2$, the energy window in which 
elastic tunneling is allowed diminishes so that the NDR occurs. 
The other origin is the potential $V_C(i)$ in Eq. (\ref{eq:uni_E}). As discussed below, 
this results in a tilting of the band in the central part, which reduces the bandwidth  
for the elastic transport effectively (see the inset of Fig. \ref{fig:fig3}). 
In (i), both factors give the NDR. In (ii), the potential gradient leads to the NDR, 
although the effect of the finite 
electrode bandwidth is absent because of the WBL. Since there is no electric field in the 
central part in (iii) and (iv), the NDR appears only when the electrode bandwidth is finite, 
which is basically the same as that in the resonant level model\cite{Baldea_PRB10}. 
These results are summarized in Table II.

\begin{figure}
\begin{center}
\includegraphics[height=6.8cm]{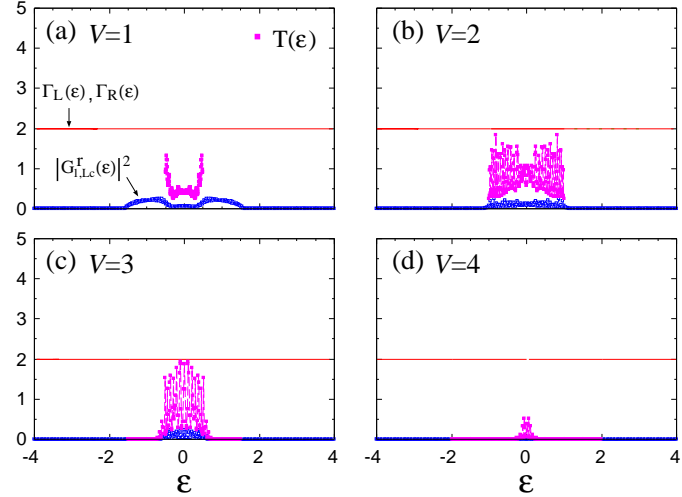}
\end{center}
\caption{(Color online) Same plot as in Fig. \ref{fig:fig3} in the case of (ii).}
\label{fig:fig4}
\end{figure}

Next, we discuss the above origins of the NDR from the energy dependence of the integrand in 
Eq. (\ref{eq:J}). 
In Fig. \ref{fig:fig3}, we show $\Gamma_L(\epsilon)$, $\Gamma_R(\epsilon)$, $|G_{1,L_C}(\epsilon)|^2$, 
and $T(\epsilon)=\Gamma_L(\epsilon)\Gamma_R(\epsilon)|G_{1,L_C}(\epsilon)|^2$ as a function of $\epsilon$ 
for several values of $V$ in the case of (i). $T(\epsilon)$ is plotted only in the region 
$\mu_R=-V/2<\epsilon <\mu_L=V/2$. The functions $\Gamma_L(\epsilon)$ and $\Gamma_R(\epsilon)$ are 
identical for $V=0$. When we increase $V$, the bands of the left and right electrodes shift in 
opposite directions, as shown in the inset of Fig. \ref{fig:fig3}. Accordingly, the region 
$\epsilon_L<\epsilon <\epsilon_R$ in which $\Gamma_L(\epsilon)\Gamma_R(\epsilon)$ is finite decreases, 
where $\epsilon_L=-2t_E+\mu_L$ ($\epsilon_R=2t_E+\mu_R$) is the energy of the band bottom (top) of 
the left (right) electrode. 
When $V$ is small, $|G_{1,L_C}(\epsilon)|^2$ is suppressed at around 
$\epsilon=0$ owing to the charge gap. This feature gradually disappears with increasing $V$. 
$|G_{1,L_C}(\epsilon)|^2$ has two peaks at $\epsilon=\pm \epsilon_{B}$ with 
$\epsilon_B\sim W/2-V/2$ and becomes vanishingly small for $\epsilon<-\epsilon_B$ 
and $\epsilon>\epsilon_B$, where $W\simeq 4$. 
We can regard $2\epsilon_B$ as the effective bandwidth for elastic transport because 
the band is tilted 
by the potential gradient. This effective bandwidth shrinks with increasing $V$. 
For $V<2$, the regions $\epsilon_L<\epsilon <\epsilon_R$ and $-\epsilon_B<\epsilon <\epsilon_B$ 
are outside the domain of integration in Eq. (\ref{eq:J}). In this case, both 
$\Gamma_L(\epsilon)\Gamma_R(\epsilon)$ and $|G_{1,L_C}(\epsilon)|^2$ contribute to $J$, 
so that $J$ increases with $V$ [Figs. \ref{fig:fig3}(a) and \ref{fig:fig3}(b)]. 
However, for $V>2$, these regions come inside the domain $\mu_R<\epsilon <\mu_L$, which results 
in the NDR [Figs. \ref{fig:fig3}(c) and \ref{fig:fig3}(d)]. The onset of the NDR by the finite 
electrode bandwidth is given by $V=W_E/2$ at which we have $\epsilon_L=\mu_R$ ($\epsilon_R=\mu_L$), 
whereas that by the potential gradient is given by $V=W/2$ because $\epsilon_B=\mu_L$ 
($-\epsilon_B=\mu_R$) holds. 

Figure \ref{fig:fig4} shows the same quantities as in Fig. \ref{fig:fig3} in the case of (ii). 
Because of the WBL, $\Gamma_L(\epsilon)$ and $\Gamma_R(\epsilon)$ are independent of $\epsilon$. 
This indicates that the energy window for the transport, $\epsilon_L<\epsilon<\epsilon_R$, 
is essentially infinite. Although $|G_{1,L_C}(\epsilon)|^2$ has no sharp peak, its $\epsilon$ 
dependence shows that the effective bandwidth $2\epsilon_B$ decreases with increasing 
$V$ as in (i). Therefore, the NDR occurs even when we 
use the WBL. For $V=1$, $T(\epsilon)$ for $\mu_R<\epsilon<\mu_L$ is similar to that in (i), 
so that the WBL becomes a good approximation for a small $V$\cite{Baldea_PRB10}. In fact, 
the $J$-$V$ curves in (i) and (ii) are quantitatively the same for $V<1.5$, as 
shown in Fig. \ref{fig:fig2}(b). When we increase $V$, their difference in $T(\epsilon)$ 
becomes larger, so that the $J$-$V$ curve in (ii) deviates from that in (i) for 
$V>1.5$. In (iv), the saturation of $J$ is understood from the WBL and $V_C(i)$ in 
Eq. (\ref{eq:no_E}). Since, in this case, $|G_{1,L_C}(\epsilon)|^2$ and 
$\Gamma_L(\epsilon)\Gamma_R(\epsilon)$ do not depend on $V$ and $\epsilon$, 
respectively, $J$ becomes constant for $V>W$. 

\begin{figure}
\begin{center}
\includegraphics[height=3.8cm]{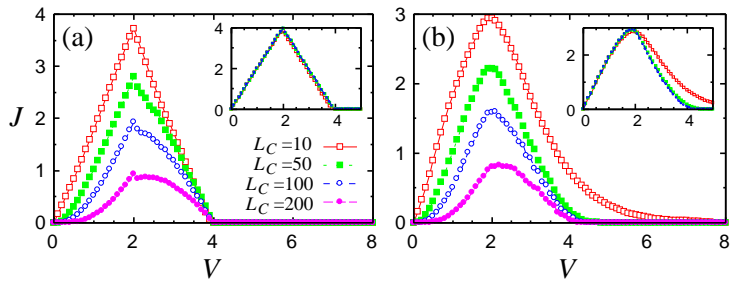}
\end{center}
\caption{(Color online) (a) Size dependence of the $J$-$V$ curve for $\delta t=0.05$ and 
$t_E=\tau=1$ in the case of (i). (b) Same plot in the case of (ii). The inset shows the 
results for $\delta t=0$.}
\label{fig:fig5}
\end{figure}

In Figs. \ref{fig:fig5}(a) and \ref{fig:fig5}(b), we show the $L_C$ dependences of the 
$J$-$V$ curve in (i) and (ii) for $t_E=\tau=1$. 
When $L_C$ is small, $J$ increases almost linearly for small values of $V$. 
The tunneling is elastic even when the charge gap exists when the correlation length 
$\xi =W/\Delta\simeq 20$ is larger than $L_C$\cite{Tanaka_PRB11}.  If we use different 
values of $\tau$ and 
$t_E$, stepwise structures in the $J$-$V$ curve become prominent, which come from the 
discreteness of the energy spectrum of the central part\cite{Tanaka_PRB11}. 
As $L_C$ increases, the $J$-$V$ curve gradually changes into the LZ-type behavior. 
When $L_C>\xi$, the deformation of the wave function in the central part plays an important 
role in the breakdown mechanism\cite{Tanaka_PRB11}. Since the threshold for the breakdown 
is determined by the electric field, 
$E_{\rm th}\propto \Delta^2/W$, $V_{th}$ is proportional to $L_C$. Therefore, 
$J$ decreases with increasing $L_C$. 
The inset shows the results for $\delta t=0$, 
which indicates that the size effect is very small when the charge gap is absent. 
For both the metal and the band insulator, we have $J=0$ for $V>4$ in the case of (i) 
since the overlap between the left and right electrodes disappears [Fig. \ref{fig:fig3}(d)]. 
In (ii), $J$ is finite even for $V>4$ owing to the WBL, although it readily approaches zero 
because the effective bandwidth of the central part vanishes at $V=4$ 
[Fig. \ref{fig:fig4}(d)].  

\begin{figure}
\begin{center}
\includegraphics[height=5.0cm]{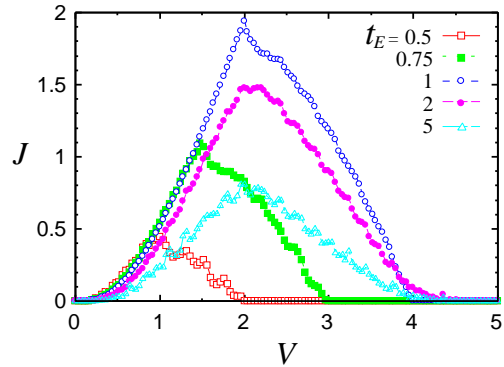}
\end{center}
\caption{(Color online) $J$-$V$ curves for several values of $t_E$ with $\delta t=0.05$, 
$\tau=1$, and $L_C=100$ in the case of (i).}
\label{fig:fig6}
\end{figure}

In Fig. \ref{fig:fig6}, we show the $J$-$V$ curves for different values of $t_E$ 
in (i) with $\delta t=0.05$, $\tau=1$, and $L_C=100$. For $t_E<1$, the NDR occurs 
at $V=W_E/2$. This is because for $W_E<W$, the threshold for the NDR by the finite 
electrode bandwidth is smaller than that by the potential gradient. On the other hand, 
the NDR sets in at $V=W/2$ for $t_E>1$ where $W_E>W$ holds. The onset of the NDR is 
determined by the smaller values of $W/2$ and $W_E/2$. 

\section{Summary}
We have investigated the $J$-$V$ characteristics of the one-dimensional band 
insulator attached to electrodes. We have shown that a linear potential gradient 
and a finite electrode bandwidth cause the NDR, the onsets of which are 
determined by $V=W/2$ and $V=W_E/2$, respectively. The former effect tilts the energy 
band of the insulator, so that the NDR results from the shrinkage of the effective bandwidth 
for elastic transport. Since this mechanism is independent of the electrode degrees 
of freedom, the NDR occurs even if we use the WBL in contrast to the noninteracting resonant 
level model. 

\begin{acknowledgments}
This work was supported by a Grant-in-Aid for Young Scientists (B) (Grant No. 
12019365) from the Ministry of Education, Culture, Sports, Science and 
Technology of Japan.
\end{acknowledgments}

\end{document}